\documentclass[aps,pra,superscriptaddress,preprint,amsmath,amssymb,showpacs]{revtex4-2}
\usepackage[russian,english]{babel}
\usepackage[utf8]{inputenc}
\usepackage{amssymb,bm}
\usepackage{graphicx}
\usepackage{amsmath}
\usepackage{color}
\usepackage{comment}
\usepackage[colorlinks=true,citecolor=blue,urlcolor=blue]{hyperref}
\begin{document}
 \begin{titlepage}

\title{The $O(2,1)$ algebra and two-dimension electron Green's function in the field of magnetic monopole}

\author{P. S. Sidorov}\email{pavel.sidorov@metalab.ifmo.ru}
\affiliation{School of Physics and Engineering, ITMO University, 197101 St. Petersburg, Russia}
\author{N. A. Vlasov}\email{nikolai.vlasov@metalab.ifmo.ru}
\affiliation{School of Physics and Engineering, ITMO University, 197101 St. Petersburg, Russia}
\author{I. S. Terekhov}\email{i.s.terekhov@gmail.com}
\affiliation{School of Physics and Engineering, ITMO University, 197101 St. Petersburg, Russia}
\affiliation{Budker Institute of Nuclear Physics of SB RAS, 630090 Novosibirsk, Russia}
\author{A. I. Milstein}\email{a.i.milstein@inp.nsk.su}
\affiliation{Budker Institute of Nuclear Physics of SB RAS, 630090 Novosibirsk, Russia}\affiliation{Novosibirsk State University, 630090 Novosibirsk, Russia}

%
%
%

\date{\today}

\begin{abstract}
Using the operator method and properties of $O(2,1)$ algebra, the integral representation for the two-dimensional Green's function  of an electron in the field of a magnetic monopole is found. This representation  is valid in all complex plane of the electron energy.
\end{abstract}

\maketitle
 \end{titlepage}
\section{Introduction}

The motion of a charged particle in the field of a magnetic monopole is interesting because of  its rich topological  content, first uncovered by Dirac \cite{Dirac}. The key result of that work was the quantization of electric charge. Despite the fact that elementary particles with magnetic charge have not been found yet \cite{Mit}, interest in this problem still persists. The reason for that is the appearance of quasiparticles with the properties of  magnetic monopoles in spin ice \cite{CasMoe, Chen, AlexSaug, Khom}. 

To study various processes  in an external field it is convenient to use the Green's function in the corresponding field. Thus, the Green's function can be represented as a series in eigenfunctions, very often integral representations for it is more convenient. For example, using the integral representation for the Green's function of an electron in a Coulomb field \cite{MilStr1982} the induced charge density was calculated exactly in the field \cite{MilStr1983}. Using integral representations for the Green's function of an electron in the field of Coulomb impurity, the induced charge in graphene and dichalcogenides were calculated  \cite{TMKS2008, IT2024}.  The integral representation for the electron Green's function in the field of narrow, infinitely long solenoid was found  in Ref. \cite{BorBos}. The induced current and the Aharonov-Bohm effect in the field  in graphene was considered in Ref.~\cite{JMPT2009}.

In this paper, the integral representation for the Green's function of a nonrelativistic charged particle moving in two spatial dimensions in the field of a magnetic monopole is obtained. At first glance, this problem appears to be a model one, since magnetic monopoles have not been found  yet. However, it is possible to study the scattering and polarization effects in the field of a monopole using the following experimental setup. Let us direct the  $z$-axis along the axis of the solenoid having length $L$. One end of the solenoid is located at $z=0$, the other one at $z=-L$. The solenoid radius $a$ obeys the condition $a\ll L$. Then, in the   plane $z=0$ at distances $a\ll\rho\ll L$, the vector potential and magnetic field take the form:
\begin{align}
	& \bm A(\bm r)=\frac{\Phi}{4\pi}\frac{[\bm\nu\times\bm\rho]}{\rho^2}\,,\quad
	\bm B(\bm r)=  \frac{\Phi}{4\pi}\frac{\bm\rho}{\rho^3}\,,\label{AandB}
\end{align}     
where  $\bm \rho=(x,y)$, $\rho=\sqrt{x^2+y^2}$, $\Phi$ is the magnetic flux through the solenoid, $\bm\nu$ is a unit vector directed along the $z$-axis. It is seen that the vector potential and magnetic field \eqref{AandB} coincide with that of a magnetic monopole having a magnetic charge equal to $\Phi/(4\pi)$. Therefore, by placing a two-dimensional electron gas in the $z=0$ plane, we can study the motion of an electron in the field of a magnetic monopole. So, the integral representation for the Green's function obtained in the present paper can apply for investigation of various processes in two-dimensional electron gas in the field of magnetic monopole.

\section{The Green's function}
In the field of a magnetic monopole, the Green's function of a charged particle with the energy $E$,  obeys the equation
\begin{eqnarray}\label{InEq}
	\left[ \left(\bm p-\frac{e}{c}\bm A\right)^2-\dfrac{e \hbar \alpha}{c} (\bm \sigma \cdot \bm B)-k^2 \right]\hat{G}(\bm\rho, \bm\rho'|k)=2 M\hat{I}\delta(\bm\rho-\bm\rho')\,,
\end{eqnarray}
where $\bm{p} =-i \hbar \left(\partial/\partial x, \partial/\partial y \right)$ is the momentum operator, $\hbar$ is the Plank constant, $M$ is the particle mass, $c$ is speed of light, $\delta(\bm \rho)$ is the Dirac $\delta$-function, $\bm \sigma=(\sigma_x,\sigma_y)$, $\sigma_{x,y}$ are the Pauli matrices, $k=\sqrt{2ME}$, $\hat{I}$ is the unit $2\times2$ matrix, the coefficient $\alpha$ takes into account the possible difference between the magnetic moment of a particle and the magnetic moment of an electron. Below we set $\hbar = c =1$, and omit the unit matrix $\hat{I}$, but imply it where necessary.

Substituting $\bm{A}$ and $\bm{B}$ \eqref{AandB} in Eq.~\eqref{InEq}, using the cylindrical coordinates, the representation 
$$\delta(\bm\rho-\bm\rho')=\frac{1}{\sqrt{\rho\rho'}}\delta(\rho-\rho')\delta(\phi-\phi'),$$ 
multiplying both sides of  Eq.~\eqref{InEq} by $\rho$, we obtain:
\begin{eqnarray}\label{Schrod2}
	\left(\frac{\partial }{\partial \rho}\rho\frac{\partial }{\partial \rho}+\frac{\hat{K}(\phi)}{\rho}+k^2\rho\right)\hat{G}(\bm r,\bm r'|k)=2M\sqrt{\frac{\rho}{\rho'}}\delta(\rho-\rho')\delta(\phi-\phi').
\end{eqnarray}
Here $\phi$ and $\phi'$ are the angles between vectors $\bm \rho$ and $\bm \rho'$ and  $x$-axis, respectively. The operator $\hat{K}$ is defined as:
\begin{eqnarray}
	\hat{K}(\phi)&=&\frac{\partial^2}{\partial\phi^2}-2i\gamma\frac{\partial}{\partial\phi}-\gamma^2
+\frac{\gamma \alpha(\bm \sigma\cdot\bm \rho)}{\rho},\label{K}
\end{eqnarray}
where  $\gamma=e\Phi/(4\pi)$. Following Ref. \cite{MilStr1982}, we use the Schwinger parametrization, and present the Green's function $\hat{G}$ in the form:
\begin{eqnarray}\label{GreenRepr}
\hat{G}_{+}(\bm \rho,\bm \rho'|k)=- 2i\,M\int\limits_0^\infty ds\exp\left\{ i s\left[k^2 \rho+\frac{\partial }{\partial \rho}\rho\frac{\partial }{\partial \rho}+\frac{\hat{K}(\phi)}{\rho} \right]\!\right\}\!\!\sqrt{\frac{\rho}{\rho'}}\delta(\rho-\rho')\delta(\phi-\phi').
\end{eqnarray}
The function $\hat{G}_{+}$ is the analytic function in the upper half-plane of the  complex  variable $k$. It is easy to find the eigenfunctions of the operator $\hat{K}$ and construct  from them the projectors which obey the equations:
\begin{eqnarray}\label{proj eq}
	&&\hat{K}(\phi)P_\lambda(\phi,\phi')=-\lambda^2P_\lambda(\phi,\phi'),\\ 
	&&\sum \limits_{\lambda}P_{\lambda}(\phi, \phi')=\sum_{m=-\infty}^{\infty}e^{im(\phi-\phi')}=\delta(\phi-\phi'),\label{DeltaDecomp}\\
	&&\int\limits_0^{2\pi}d\phi'P_{\lambda}(\phi, \phi')P_{\lambda}(\phi', \phi'')=P_{\lambda}(\phi, \phi'').
\end{eqnarray} 
There are two sets of eigenvalues  $\lambda_{1}$ and  $\lambda_{2}$:
\begin{eqnarray}\label{Eigenvalues}
	\lambda_{1}&=&\sqrt{\nu^2+\varkappa+\frac{1}{4}}\,,\,\,\lambda_{2}=\sqrt{\nu^2-\varkappa+\frac{1}{4}}\,,\\
	\varkappa&=&\sqrt{\alpha^2\gamma^2+\nu^2}\,,\,\,\nu=m-\gamma+1/2.
\end{eqnarray}
The sum over $\lambda$ means the summation over $\lambda_{1,2}$ and $m.$ Note that  $\lambda_1$ is always positive, while $\lambda_2$ can be real and positive or purely imaginary number. In the latter case  $\lambda_2= \pm i|\lambda_2|$ and both signs should be taken into account. The relative contributions are defined by the additional boundary condition, see below. 

The explicit forms of the projectors are 
\begin{eqnarray}{\label{Projectors}}
	P_{\lambda_{1,2}}(\phi,\phi')&=&\frac{e^{im(\phi-\phi')}}{4\pi \varkappa}\left(
	\begin{array}{cc}
		\varkappa \mp\nu&\mp\gamma \alpha\,e^{-i\phi'}\\
		\mp\gamma \alpha\,e^{i\phi}&(\varkappa \pm\nu)e^{i(\phi-\phi')}
	\end{array}
	\right)\,,
\end{eqnarray}
where the upper sign corresponds to $\lambda_1$, and the lower one to $\lambda_2$.

Substituting the decomposition \eqref{DeltaDecomp} to Eq.~\eqref{GreenRepr}, we obtain:
\begin{eqnarray}\label{GreenRepr2}
	\hat{G}_{+}(\bm \rho,\bm \rho'|k)=-2iM\sum_{\lambda}P_\lambda(\phi,\phi')\int\limits_0^\infty dse^{i s\left(k^2 \hat{T}_3-2\hat{T}_1 \right)}\sqrt{\frac{\rho}{\rho'}}\delta(\rho-\rho'),
\end{eqnarray}
where the operators $\hat{T}_1$, $\hat{T}_3$, together with the operator $\hat{T}_2$, 
\begin{eqnarray}
	\hat{T}_1=-\frac{1}{2}\left(\partial_\rho \rho\partial_\rho-\frac{\lambda^2}{\rho}\right), \quad	\hat{T}_3=\rho, \quad \hat{T}_2=-i\left(\rho\,\partial_\rho+\frac{1}{2}\right),
\end{eqnarray}
are the generators of the $O(2,1)$ algebra:
\begin{eqnarray}\label{Algebra}
	[\hat{T}_1,\hat{T}_2]=-i\hat{T}_1\,,\quad [\hat{T}_1,\hat{T}_3]=-i\hat{T}_2\,,\quad [\hat{T}_2,\hat{T}_3]=-i\hat{T}_3\,.
\end{eqnarray}
So, to find the integral representation we should calculate the action of operator $e^{i s\left(k^2 \hat{T}_3-2\hat{T}_1 \right)}$ on $\delta(\rho-\rho')$.
Some examples of applying the $O(2,1)$ algebra in a Coulomb field are presented in Refs.~\cite{Hambu,DmitRum,BaierMilst1977,LeeMil}. In Ref. \cite{MilStr1982} the Green's function for an electron in the Coulomb field was found using the $O(2,1)$ algebra.  In Ref.~\cite{JMPT2009} the Green's function was found for an electron in graphene in the field of \textcolor{red}{a} narrow infinitely long solenoid.  Using the results of Ref.~\cite{MilStr1982,JMPT2009} we obtain for positive $\lambda_1$ and $\lambda_2$: 
\begin{eqnarray}\label{E_action}
&&e^{i s\left(k^2 \hat{T}_3-2\hat{T}_1 \right)}\sqrt{\frac{\rho}{\rho'}}\delta(\rho-\rho')=\frac{kJ_{2\lambda}\left(y\right)}{\sinh (ks)}e^{ ik (\rho+\rho')\coth(ks)-i\pi\lambda}\,,\\
&&y=\frac{2k\sqrt{\rho\rho'}}{\sinh(k s)}\,.\label{y}
\end{eqnarray}
where $J_a(x)$ is the Bessel function. 

To perform the  calculation for the  imaginary $\lambda_2$, it is necessary to consider the "fall to the center"\, phenomenon, i. e., the behavior of the wave function $\psi_{\lambda_2}(\rho,\phi)$ at small distances $k\rho\ll1$.  At such distances, the Schr\"odinger equation for $\psi_{\lambda_2}(\rho,\phi)$ transforms to
\begin{eqnarray}
	&&\hat{T}_1\psi_{\lambda_2}(\rho,\phi)=0\,,\\
	&&\psi_{\lambda_2}(\rho,\phi)= {\cal V}_{\lambda_2}(\phi)f(\rho),\\
	&&\hat K(\phi){\cal V}_{\lambda_2}(\phi)=-\lambda_2^2{\cal V}_{\lambda_2}(\phi).
\end{eqnarray}
The general solution for the function $f(\rho)$ is
\begin{eqnarray}\label{WaveFuncAsymp}
	f(\rho)=C\rho^{\lambda_2}+D\rho^{-\lambda_2},
\end{eqnarray}
where $C$ and $D$ are some constants. Since the wave function should be normalized we set $D=0$ for positive $\lambda_2$. For imaginary $\lambda_2$ there is no such restriction. So, we have two independent solutions for the wave function with the asymptotics $\rho^{\pm i|\lambda_2|}$ at small $\rho$. This leads to ambiguity when constructing the Green's function. To avoid the ambiguity  it is necessary to introduce the additional boundary conditions at small distances, see below. For imaginary $\lambda_2^2<0$  the expression \eqref{E_action} should be modified.  We represent the $\delta$-function as follows
 \begin{eqnarray}\label{DeltaDecomp1}
 	\sqrt{\frac{\rho}{\rho'}}\delta(\rho-\rho')=\alpha \rho^{i|\lambda_2|}\int\limits_{-i\infty}^{i\infty}d\sigma e^{\sigma\rho}g_1(\sigma,\rho')+\beta \rho^{-i|\lambda_2|}\int\limits_{-i\infty}^{i\infty}d\sigma e^{\sigma\rho}g_2(\sigma,\rho'),
 \end{eqnarray}
 where $g_{1,2}(\sigma,\rho')=(\rho')^{\mp i|\lambda_2|}e^{-\sigma\rho'}\,,$ $\alpha+\beta=1$.  Then we use the results of Refs. \cite{MilStr1982,JMPT2009} for the first and second terms in the right-hand side  in Eq. \eqref{DeltaDecomp1} and obtain
\begin{eqnarray}\label{E_action1}
e^{i s\left(k^2 \hat{T}_3-2\hat{T}_1 \right)}\sqrt{\frac{\rho}{\rho'}}\delta(\rho-\rho')=\frac{k e^{ik (\rho+\rho')\coth(ks)}}{\sinh (ks)}
\left(\alpha e^{\pi|\lambda_2|}
J_{2i|\lambda_2|}\left(y\right)+\beta e^{-\pi|\lambda_2|}
J_{-2i|\lambda_2|}\left(y\right)\right).
\end{eqnarray}
Substituting Eqs.~\eqref{E_action} and \eqref{E_action1} to Eq.~\eqref{GreenRepr2} we obtain the integral representation for the Green's function:
\begin{eqnarray}\label{G}
	\hat{G}_{+}(\bm \rho,\bm \rho'|k)=\hat{G}_{+}^{(R)}(\bm \rho,\bm \rho'|k)+\hat{G}_{+}^{(I)}(\bm \rho,\bm \rho'|k)\,,
\end{eqnarray}
\begin{eqnarray}
	\hat{G}_{+}^{(R)}(\bm \rho,\bm \rho'|k)=-2iM\sum_{\lambda>0 }P_\lambda(\phi,\phi')\int\limits_0^\infty \frac{k ds}{\sinh (ks)}e^{ik (\rho+\rho')\coth(ks)-i\pi\lambda}
	J_{2\lambda}\left(y\right)\,,
\end{eqnarray}
\begin{eqnarray}\label{G_Irreg}
	\hat{G}_{+}^{(I)}(\bm \rho,\bm \rho'|k)&=&- 2iM\sum_{\lambda=i|\lambda| }P_\lambda(\phi,\phi')\int\limits_0^\infty \frac{k ds}{\sinh (ks)}e^{ik (\rho+\rho')\coth(ks)}\nonumber\\
	&\times&	\left(\alpha_m e^{\pi|\lambda|}
	J_{2i|\lambda|}\left(y\right)+\beta_m e^{-\pi|\lambda|}
	J_{-2i|\lambda|}\left(y\right)\right),
\end{eqnarray}
where  $\alpha_m+\beta_m=1$, and $y$ is defined in Eq.~\eqref{y}. To find  $\alpha_m$ and $\beta_m$ separately we should take into account that in our setup there is no   "fall to the centre"\, phenomenon, since the magnetic field $\bm{B}$ and \textcolor{red}{the} vector potential $\bm{A}$ are not singular at small distances $\rho\sim a$ ($a$ is the solenoid radius). Thus, the coefficients $C$ and $D$ in Eq.~\eqref{WaveFuncAsymp} are fixed by the behavior of the wave function at the distances $\rho\sim a$. So, the relation between the coefficients $\alpha_m$ and $\beta_m$ depends on the experimental conditions. 

For example, let us consider the case when the region $\rho<R$ is forbidden for electron penetration. So that, the electron  radial current vanishes at $\rho=R$. Then for $kR\ll1$ we obtain: 
%
%
\begin{eqnarray}
	\alpha_m=\beta_m=1/2.
\end{eqnarray}   
It is easy to check that the  Green's funciton $\hat{G}_{-}(\bm \rho,\bm \rho|k)$, which is  analytic function in the lower half-plane of a complex variable $k$, can be obtained from $\hat{G}_{+}(\bm \rho,\bm \rho|k)$ by integration over the variable $s$ from $-\infty$ to $0$.

\section{Conclusion}

The integral representation for the two-dimensional Green's function  of an electron in the field of a magnetic monopole is obtained. The magnetic field of the monopole is generated by a semi-infinite solenoid. The integral representation is valid in all complex plane of the energy $E$. It is shown that  the uncertainty in the Green's function corresponds to the uncertainty of the electron  wave function at small distances. The latter is fixed by the physical condition of the given experiment.

\section*{Acknowledgement}
The work of I.S.T. was financially supported by the ITMO Fellowship Program. Data sharing not applicable to this article as no datasets were generated or analysed during the current study. The authors declare no conflicts of interest.

\end{document}